\documentclass{acm_proc_article-sp}
\usepackage{amsmath}
\usepackage{tabularx}
\usepackage{algorithm}
\usepackage{algorithmic}
\usepackage{mdwlist}
\usepackage{graphicx}
\usepackage{balance}  
\usepackage[colorlinks,linkcolor=red,citecolor=blue,urlcolor=blue,draft]{hyperref}
\usepackage{url}
\usepackage{tikz}
\usetikzlibrary{positioning,patterns,shapes}
\usepackage{pgfplots}

\tikzset{
  block/.style={
    fill=white,
    draw=black,
    thick,
    rectangle,
    rounded corners,
    text centered}
}

\tikzset{
  decision/.style={
    fill=white,
    draw=black,
    thick,
    diamond,
    text centered}
}

\tikzset{
  influence/.style={
    fill=black!30,
    draw=black,
    thick,
    rectangle,
    rounded corners,
    text centered}
}

\input{Definitions}

\begin{document}

\title{Modeling Attractiveness and Multiple Clicks in Sponsored Search
  Results}

\numberofauthors{3}
\author{
 \alignauthor
 Dinesh Govindaraj\\
 \affaddr{Microsoft Bing Ads}\\
 \affaddr{Bangalore}\\
 \email{digovind@microsoft.com}
 \alignauthor Tao Wang\raisebox{9pt}{$\ast$} \\
 \affaddr{Purdue University}\\
 \affaddr{West Lafayette, IN}\\
 \email{taowang@purdue.edu}
 \and  
 \alignauthor S V N Vishwanathan\titlenote{Work done when visiting Microsoft Bing Ads, Bangalore.}\\
 \affaddr{Purdue University}\\
 \affaddr{West Lafayette, IN}\\
 \email{vishy@stat.purdue.edu}
}

\maketitle

\begin{abstract}
  Click models are an important tool for leveraging user feedback, and
  are used by commercial search engines for surfacing relevant search
  results. However, existing click models are lacking in two
  aspects. First, they do not share information across search results
  when computing attractiveness. Second, they assume that users interact
  with the search results sequentially.  Based on our analysis of the
  click logs of a commercial search engine, we observe that the
  sequential scan assumption does not always hold, especially for
  sponsored search results. To overcome the above two limitations, we
  propose a new click model. Our key insight is that sharing information
  across search results helps in identifying important words or
  key-phrases which can then be used to accurately compute
  attractiveness of a search result. Furthermore, we argue that the
  click probability of a position as well as its attractiveness changes
  during a user session and depends on the user's past click
  experience. Our model seamlessly incorporates the effect of
  externalities (quality of other search results displayed in response
  to a user query), user fatigue, as well as pre and post-click
  relevance of a sponsored search result. We propose an efficient
  one-pass inference scheme and empirically evaluate the performance of
  our model via extensive experiments using the click logs of a large
  commercial search engine.
\end{abstract}



\keywords{Click Models, User behavior, Query log, Click and Browsing
  data analysis, Sponsored Search, Optimization,
  Ranking} 

\section{Introduction}
\label{sec:Introduction}

When a user submits a query, search engines return \emph{organic search
  results} as a list of ranked URLs. Sometimes URLs linked to landing
pages of ads (sponsored search results) are also displayed along with
the organic search results. Sponsored search URLs displayed above the
organic search results are called \emph{mainline ads}, while those
displayed on the side are called \emph{sidebar ads}. Search engines
generate revenue when users click on ads.  In this paper we are
interested in modeling how users interact with and click on sponsored
search results. In particular, we focus on mainline ads since a majority
of user clicks and hence revenue can be attributed to these ads.

Mining click-through logs is an important component of the never ending
quest of commercial search engines to surface the most \emph{relevant}
(sponsored) search results in response to an user query. To understand
this, consider the following scenario: Suppose documents $d$ and $d'$
are displayed in response to a query $q$. If the number of clicks that
$d$ receives is disproportionately high compared to $d'$, one can
reasonably conclude that $d$ is more relevant than $d'$ for $q$. While
this approach of learning from the ``wisdom of the crowd'' is cheaper
than employing a human labeler, it is not without its fair share of
difficulties. For instance, it is well known that user clicks display a
position bias, that is, documents at higher ranked positions are more
likely to be clicked than lower ranked documents. There is a rich body
of research which tries to infer an unbiased estimate of the document
relevance from click-through logs by explicitly modeling user behavior
using click models.

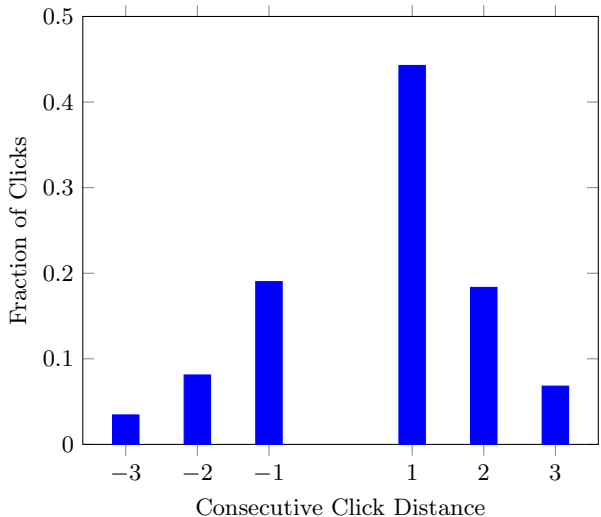
\begin{figure}[H]
  \begin{center}
    \begin{tikzpicture}[scale=1.0] 
      \begin{axis}[ybar,
        xlabel=Consecutive Click Distance, xlabel near ticks,
        ylabel=Fraction of Clicks, ylabel near ticks,
        ymin=0.0, ymax=0.5,
        xtick=data,]
        \addplot[draw=blue,fill=blue] table[x=ClickDist,y=FracClicks] {click_dist.txt};
      \end{axis}
    \end{tikzpicture}
  \end{center}
  \caption{If the user clicks on position $i$ followed by position $j$,
    then the click distance is defined as $j-i$. Since at most 4
    mainline ads are shown, the click distance lies in the range
    $[-3,3]$. We plot a normalized histogram of the distance between
    consecutive clicks.}
  \label{fig:click-dist}
\end{figure}

Almost all existing click models are designed for organic search, and
make the simplifying assumption that users interact with the search
results sequentially \cite{Joachims02,JoaGraPanHemetal05}. In other
words, they assume that a user examines an URL at position $i+1$ only
after examining the URLs at positions $1,\ldots, i$. Furthermore, since
an overwhelming majority of user sessions end after one click, many
models focus exclusively on this scenario. However, based on the
analysis of the click logs of a commercial search engine, we observed
that user interaction with sponsored search results do not confirm to
these assumptions. Approximately 10\% of the user sessions with clicks
contain more than one click. Furthermore, in approximately 30\% of
multi-click sessions at least one pair of clicks is in reverse order
(\eg, a click at position 3 followed by a click at position 1). This
situation is depicted graphically in Figure~\ref{fig:click-dist}.

We conjecture that users scan all the sponsored search results before
they begin clicking. This is because sponsored search results are shown
in a small block at the beginning of the page and usually occupy only 4
to 8 lines.  Perhaps, eye tracking studies \cite{GraJoaGay04} which are often used to support the
linear scan assumption in organic search may not apply to sponsored
search\footnote{Validating our conjecture using carefully designed user
  studies is an interesting research direction in its own right, but is
  unfortunately beyond the scope of the current work}. Therefore,
meaningful click models for sponsored search need to model reverse
clicks.

Unlike organic search, where short snippets of the landing page of a
search result is displayed, sponsored search results are designed to
grab attention. Towards this end, they employ titles that are short,
catchy, and some words (which match query terms) are displayed in bold
font. An sponsored search result might look very attractive to an user,
but after clicking on it she might realize that the landing page does
not contain the information that she wants. In other words, not only the
\emph{attractiveness} but also the \emph{post-click relevance} of a
sponsored search result are important factors for user satisfaction in
sponsored search.

Somewhat surprisingly, most existing click models either do not
distinguish between attractiveness and post-click relevance or use very
simple methods of estimating attractiveness.  However, consider the
following: the presence of the words ``bonded and insured'' in the title
can make a sponsored search result more attractive for the query
``plumber''. If we can share such information across all sponsored
search results displayed in response to the query ``plumber'', we can
accurately estimate attractiveness.

In this paper, we present a novel statistical click model which models
reverse clicks, and incorporates a new method for estimating 
attractiveness. Like in previous work
\cite{RicDomRag07,SriBasWanPre10,CheYan12} we also assume a separable
model, that is, the user propensity for clicking on a search result is a
product of the probability of examining that position and the
attractiveness of the document displayed at that position. However, our
key insight is that the examination probability of a position as well as
the attractiveness of a position changes over time and depends on the
user's past click experience.

Our contributions can be summarized as follows: We propose a new click
model for user interaction with sponsored search results. Our model can
handle multiple click sessions as well as user sessions where the order
of clicks is reversed. Furthermore, we present a new method for
estimating attractiveness which shares information across multiple
sponsored search results displayed in response to a user query.  Our
model is flexible and can easily incorporate user fatigue, pre-click and
post-click relevance of an ad, and externalities. We derive an efficient
one-pass inference mechanism for estimating parameters. Finally, we show
that our model comprehensively outperforms existing click models in
large scale empirical evaluation using real-life data from a large
commercial search engine.

\section{Click Models: A Brief Review}
\label{sec:Background}

In what follows we will use the terms documents, ads, urls, and
sponsored search results interchangeably. Almost all click prediction
models make the following separability assumptions: (1) A document is
clicked if and only if it is examined (or viewed); (2) The probability
that a document is clicked is independent of its position given that it
was viewed; (3) The probability that a document is viewed is independent
of the document given the position, and is independent of the other ads
presented.

\paragraph{Examination Hypothesis}
\label{sec:ExamHypoth}

Let $E_i$ (resp.\ $C_i$) be a random variable which is equal to 1 if a
document $d_i$ at position $i$ was examined (resp.\ clicked). Based on the
above assumptions one can write:
\begin{align}
  \nonumber
  P(C_i=1\,|\, d_i) &= P(C_i=1 \cap E_i=1 \,|\, d_i)\\
  & = \underbrace{P(C_i=1 \,|\, d_i,
    E_i=1)}_{\text{relevance}}\underbrace{P(E_i=1)}_{\text{position
      bias}}.
  \label{eq:exam-hypo}
\end{align}
In other words, the probability of a document being clicked can be
factored into the product of the relevance of the document and the
position bias. This is the so-called examination hypothesis
\cite{RicDomRag07} or separability assumption \cite{AggGoeMot06}. 

\paragraph{Cascade Model and Extensions}
\label{sec:CascadeModels}

The cascade model \cite{CraZoeTayRam08} assumes that users scan the
documents from top to bottom without skipping, that is, users examine
the ads sequentially:
\begin{align}
  \label{eq:cascade-hypo}
  P(E_i=1|E_{i-1}=1) = 1 \text{ and } P(E_i=1|E_{i-1}=0) = 0. 
\end{align}
The cascade model \cite{CraZoeTayRam08} further assumes that users
examine documents until they find an appropriate one, and then abandon
the search after the first click:
\begin{align}
  \label{eq:cascade-model}
  P(E_i=1|E_{i-1}=1,C_{i-1}) = 1-C_{i-1}. 
\end{align}
The basic cascade model can only deal with query sessions with a single
click. The dependent click model (DCM) \cite{GuoLiuWan09} extends this
to multi-click sessions, by modifying \eqref{eq:cascade-model} to
\begin{align}
  \label{eq:dcm-model}
  P(E_i=1|E_{i-1}=1, C_{i-1}) = \rbr{\lambda_{i-1}}^{C_{i-1}}. 
\end{align}
Here $\lambda_{i}$ is estimated from the empirical probability that a
browsing session did not end after a click at position $i$. 

The user browsing model (UBM) \cite{DupPiw08} extends the DCM by
assuming that the examination probability depends on the last clicked
position in the same query session. In other words
\begin{align}
  \label{eq:ubm}
  P(E_i=1\,|\,C_{1},\ldots, C_{i})=\beta_{r_{i}},
\end{align}
where $r_{i}$ denotes the position of the previous click, that is,
$r_{i} := \argmax_{j} I\rbr{C_{j}=1}$. The Bayesian browsing model (BBM)
\cite{LiuGuoFal09} models relevance $P(C_i=1 \,|\, d_i, E_i=1)$ as a
random variable, and uses a Bayesian approach to estimate its value.

\paragraph{Dynamic Bayesian Network Model}
\label{sec:DynamBayesNetw}

The dynamic Bayesian network (DBN) \cite{ChaZha09} model differs from
the above click models in two important ways. First, it distinguishes
between attractiveness (also called perceived relevance or pre-click
relevance in literature \cite{SriBasWanPre10}) and post-click
relevance. Second, it incorporates user fatigue. This leads to
\begin{align}
  \label{eq:dbn-model-click}
  P(C_{i-1}=1|d_{i-1}, E_{i-1}=1) & = \theta_{d_{i-1}}\\
  \label{eq:dbn-model-examine}
  P(E_{i}=1|E_{i-1}=1, C_{i-1}) & = \lambda_{i-1}
  \rbr{1-\rho_{d_{i-1}}}^{C_{i-1}}.
\end{align}
Here, $\rho_{d_{i-1}}$ (resp.\ $\theta_{d_{i-1}}$) denotes the
post-click satisfaction (resp.\ attractiveness) of document $d_{i-1}$
displayed at position $i-1$.  For simplicity, one can set
$\lambda_{i-1}=\lambda$ for all $i$ \cite{ChaZha09}, or like in the DCM
estimate $\lambda_{i-1}$ from empirical probabilities. Similarly,
$\theta_{d_{i-1}}$ is estimated (essentially) by counting the number of
times a document was displayed and the number of times it was
clicked. As we will see later in the paper, estimating the
attractiveness by using a naive-Bayes like model leads to information
sharing and hence better estimates for the attractiveness. 

Recently \cite{AshCla12} extended the DBN by learning a per-query
fatigue parameter and adding a variable to indicate if the user is
likely to interact with sponsored search results or directly skip over
them. A related model to DBN is the click chain model (CCM)
\cite{GuoLiuKanMinetal09}.

\paragraph{Temporal Hidden Click Model}
\label{sec:TemporalHiddenClick}

Some recent work \cite{XuLiuZhaMaetal12} has focused on incorporating
revisiting behaviors into click models. However, the temporal hidden
click model (THCM) proposed by the authors of \cite{XuLiuZhaMaetal12}
does not distinguish between attractiveness and post-click
relevance. Furthermore, THCM allows for only two kinds of transitions
namely
\begin{align}
  \label{eq:thcm}
  P(E_{i}=1|E_{i-1}=1) & = \alpha \\
  P(E_{i-2}=1|E_{i-1}=1) & = \gamma,
\end{align}
with $\alpha + \gamma \leq 1$. Note that the transition probabilities
do not depend on the location $i-1$. Furthermore, the probability of
examining a position decays exponentially with distance from the current
location. 

\paragraph{Effect of Externalities}
\label{sec:EffectExternalities}

Externalities \cite{GhoMah08} refer to the observation that the click on
a document also depends on the quality of other documents presented on
the same search result page. That is, the examination event will be
affected not only by the documents shown above a certain position, but
also by the documents below a certain position.

Much of the early work on click models considers the displayed documents
as being independent of each other.  Some recent work has tried to take
externalities into account. For instance, \cite{XioWanDinSheetal12}
proposed a conditional random fields based model to click prediction to
consider the externalities., \cite{XuManCan10} considers externalities
but restrict their attention to sessions with just two ads. On the other
hand, \cite{SriBasWanPre10} showed that the relevance of a document at a
position is not constant and it is affected by the clicks in other
positions. In the context of news recommendation, \cite{AhmTeoVisSmo12}
extended the basic DBN model to take into account a sub-modular
information gain score, which affects the relevance of a
document. Although a similar extension is also possible in our setting,
for the sake of brevity we will omit discussing this for our model.



\section{Model Description}
\label{sec:NewModel}
\subsection{Notation}
\label{sec:Notation}

Since we have to deal with reverse clicks, our notation is
non-standard. Given a query $q$, we assume that the user is presented
with a ranked list of $n$ documents or sponsored search results $\Dcal =
\cbr{d_{1}, \ldots, d_{n}}$, and she might choose to click on $0 \leq
\nhat \leq n$ documents. We consider this as a query session.  Let
$c_{i} \in \cbr{0, 1, \ldots, n}$ be a multinomial random variable;
$c_{i}$ for $1 \leq i \leq \nhat$ denotes the position of the $i$-th
click, and we let both $c_{0}$ and $c_{\nhat + 1}$ to be equal to
$0$. Let $\cbb_{1:i} = \rbr{c_1,c_2,\cdots,c_i}$, and $\cbb$ be a
shorthand for $\cbb_{1:\nhat+1}$.\\

\begin{figure}[thb]
  \begin{center}
    \begin{tikzpicture}[scale=1.0] 
      \node[block] at (4, 9) (start) {Start};
      \node[decision] at (4, 7) (abandon) {Abandon?};
      \node[block] at (4, 5) (select) {Select a position to click};
      \node[decision] at (4, 3) (satisfied) {Satisfied?};
      \node[block] at (4, 1) (stop) {Stop};
      \draw[->,thick] (start) -- (abandon);
      \draw[->,thick] (abandon) -- (select) node at (4.3, 5.7) {No};
      \draw[->,thick] (select) -- (satisfied);
      \draw[->,thick] (satisfied) -- (stop) node at (4.3, 1.7) {Yes};
      \draw[->,thick]  (abandon.west) -- +(-2.5,0) -- +(-2.5,-6) --
      (stop.west) node at (2.5, 7.2){Yes};
      \draw[->,thick]  (satisfied.east) -- +(2.5,0) -- +(2.5,4) --
      (abandon.east) node at (5.5, 3.2){No};
      \node[influence] at (1.5, 6.3) (prevclicks) {Prev. clicks};
      \node[influence] at (6.3, 6.3) (attr) {Attractiveness};
      \draw[->,thick] (attr) -- (select);
      \draw[->,thick] (prevclicks) -- (select);
    \end{tikzpicture}
  \end{center}
  \caption{User browsing behavior.}
  \label{fig:user-behavior}
\end{figure}
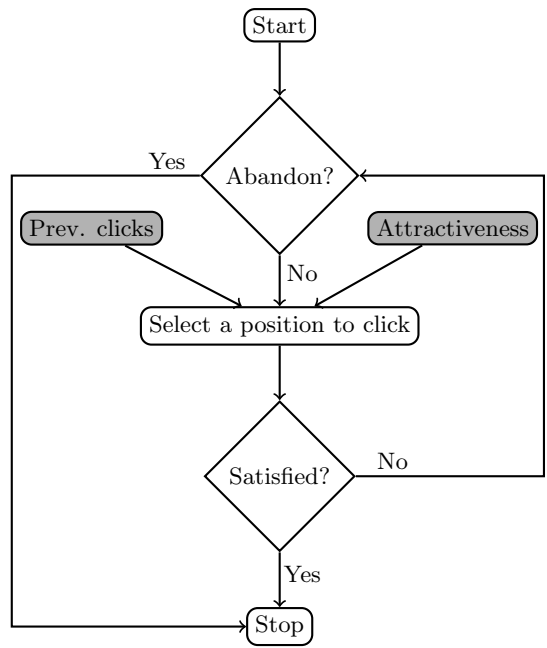

The user browsing behavior in sponsored search can be explained by
Figure~\ref{fig:user-behavior}. After clicking on a document, the user
has two choices: she can abandon the session or she can choose to return
and click on any other document that has not been clicked so
far. Users abandon a session for one of two reasons:
\begin{itemize}
\item In the \emph{abandon with satisfaction} case, the user found the
  information she was looking for in the sponsored search results. We
  capture this behavior via a Bernoulli random variable $s_i$, which
  takes on a value of $1$ when the user is satisfied after clicking the
  document $d_{c_{i-1}}$ at position $c_{i-1}$. This, in turn, depends
  on the \emph{post-click relevance} of the clicked document, which we
  denote as $\rho_{d_{c_{i-1}}}$. In other words, 
  \begin{align}
    \label{eq:probsdef}
    P\rbr{s_{i} = 1 \,|\, c_{i-1}, \Dcal} = \rho_{d_{c_{i-1}}}.
  \end{align}
  By default, we will set $P\rbr{s_{i} = 1 \,|\, c_{0}, \Dcal} = 0$.
\item In the \emph{abandon without satisfaction} case, the user may
  choose to skip ahead to the organic search results or close the
  browsing session. This behavior is captured via another Bernoulli
  random variable $t_{i}$, which depends upon the \emph{perseverance} of
  the user. We will let 
  \begin{align}
    \label{eq:probtdef}
    P\rbr{t_{i} = 1 \,|\, s_i = 0, c_{i-1}} = \eta_{i}, 
  \end{align}
  where $\eta_{i}$ denotes the probability that user will abandon the
  session after $i$ clicks.
\end{itemize}
As one can expect, the perseverance of the user decreases as the number
of clicks increase. This allows our model to capture the effect of
externalities in the following natural way: The user fatigue parameter
multiplied by the pre-click relevance gives the instantaneous pre-click
relevance of a document. If an ad appears alongside other highly
relevant ads which have already received clicks, then its instantaneous
pre-click relevance reduces. In other words, the number of previous
clicks observed is a direct measure of \emph{externalities} (relevance
of other documents).  On the other hand, when the user decides to
continue, then two factors influence the position of the next click:
\begin{itemize}
\item She will choose a position based on \emph{attractiveness} of the
  documents presented. We denote the attractiveness of the document at
  position $c_{i}$ by $a_{c_{i}}$, and assume that the attractiveness of
  the documents at different positions is independent of each
  other. Furthermore, let $\ab$ denote the $n$-dimensional vector of
  $a_{i}$ and 
  \begin{align}
    \label{eq:probadef}
    P\rbr{a_{c_{i}}=1 \,|\, \Dcal}=\theta_{d_{c_{i}}}.
  \end{align}
\item The position of the previous click influences the position of the
  next click, the so-called \emph{position bias} effect. This is
  captured using $\vb_{i}$ a $n$ dimensional random vector, whose
  distribution is given by $\gammab_{c_{i-1}}$
  \begin{align}
    \label{eq:ebdef}
    P\rbr{v_{i,j}=1 \,|\,c_{1:i-1}} & =  \gamma_{c_{i-1},j}.
  \end{align}
  Here $\gamma_{c_{i-1},j}$ denotes the probability that the user
  transitions to position $j$ after the previous click at position
  $c_{i-1}$. We allow for non-zero $\gamma_{c_{i-1}, j}$ even if $j <
  c_{i-1}$. This is different from existing click models which assume
  that users scan the list linearly, that is $c_{i} > c_{i-1}$.
\end{itemize}



\subsection{Graphical Model}
\label{sec:GraphicalModel}

\begin{figure}[htbp]
  \centering
  \begin{tikzpicture}
    \draw (-1.0, 2.5) node (cim1) [draw,circle,fill=black!20,inner sep=8pt] {};
    \node [right=0pt of cim1] {$c_{i-1}$};

    \draw (1.0, 2.5) node (ci) [draw,circle,fill=black!20,inner sep=8pt] {};
    \node [right=0pt of ci] {$c_{i}$};

    \draw (1.0, 4.5) node (ti) [draw,circle,inner sep=8pt] {};
    \node [right=0pt of ti] {$t_{i}$};

    \draw (1.0, 0.5) node (vi) [draw,circle,inner sep=8pt] {};
    \node [right=0pt of vi] {$\vb_{i}$};

    \draw (-1.0, 0.5) node (si) [draw,circle,inner sep=8pt] {};
    \node [right=0pt of si] {$s_{i}$};

    \draw[->,>=stealth] (cim1) -- (ti);
    \draw[->,>=stealth] (cim1) -- (vi);
    \draw[->,>=stealth] (ti) -- (ci);
    \draw[->,>=stealth] (cim1) -- (si);
    \draw[->,>=stealth] (si) -- (ci);
    \draw[->,>=stealth] (vi) -- (ci);

    \draw[thick] (-2.0,-0.2) rectangle (2.0,5.2);
    
    \draw (-3.0, 4.5) node (a) [draw,circle,inner sep=8pt] {};
    \node [left=0pt of a] {$\ab$};

    \draw (-3.0, 2.5) node (d) [draw,circle,fill=black!20,inner sep=8pt] {};
    \node [left=0pt of d] {$\Dcal$};

    \draw[->,>=stealth] (a) -- (cim1);
    \draw[->,>=stealth] (a) -- (ci);
    \draw[->,>=stealth] (d) -- (a);
    \draw[->,>=stealth] (d) -- (si);

    \node at (-1.,4.75) {$i=1:\nhat+1$};

    \draw (1.0, -1.0) node (vip) [draw,circle,fill=black!20,inner sep=8pt] {};
    \node [right=0pt of vip] {$\Ab^{\gamma}$};
    \draw[->,>=stealth] (vip) -- (vi);

    \draw (1.0, 6.0) node (tip) [draw,circle,fill=black!20,inner sep=8pt] {};
    \node [right=0pt of tip] {$\alphab^{\eta}$};
    \draw[->,>=stealth] (tip) -- (ti);

  \end{tikzpicture}  
  \caption{Our click prediction model represented as a Bayesian
    Network. Shaded nodes are observed. }
  \label{fig:graphical-model}
\end{figure}
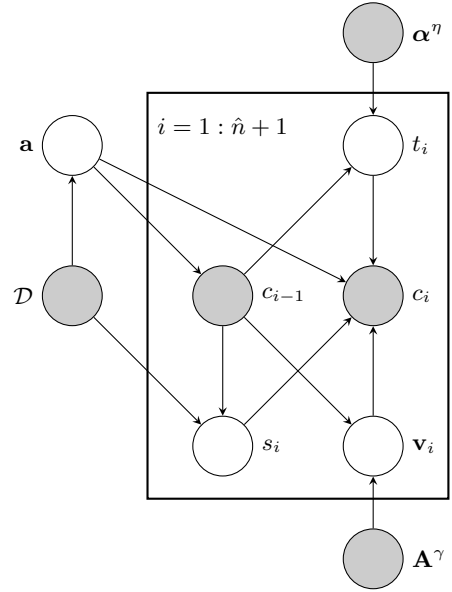

Following standard practice, we will place a $Beta$ prior on
$\eta_{i}$
\begin{align}
  \label{eq:eta-prior}
  \eta_{i} & \sim Beta(\alpha^{\eta}_{i}, \beta^{\eta}_{i})
\end{align}
and a Dirichlet prior on $\gammab_{i}$
\begin{align}
  \label{eq:gamma-prior}
  \gammab_{i} & \sim Dirichlet(\alphab^{\gamma}_{i}).
\end{align}
A Bayesian network specification of our model can be found in
Figure~\ref{fig:graphical-model}.
We denote all vectors in bold and matrices in capital and bold. If we
let $\sbb=\rbr{s_{1}, \ldots, s_{n}}$, $\tb=\rbr{t_{1}, \ldots, t_{n}}$,
$\Vb = \rbr{\vb_{1}, \ldots, \vb_{n}}$, $\alphab^{\eta} =
\rbr{(\alpha^{\eta}_1, \beta^{\eta}_1), \ldots, (\alpha^{\eta}_n,
  \beta^{\eta}_n)}$, and $\Ab^{\gamma} = \rbr{\alphab^{\gamma}_1, \ldots,
  \alphab^{\gamma}_n}$, then our click model can be described as
follows:
\begin{align}
  \label{eq:ClickModel}
  & P(\cbb, \sbb, \tb,\ab, \Vb \,|\, \alphab^{\eta},
  \Ab^{\gamma}, \Dcal) = \\
  \nonumber & P(\ab\,|\,\Dcal) \times P(\sbb\,|\,\Dcal) \times
  P(\alphab^{\eta}\,|\,\Dcal) \times
  P(\Ab^{\gamma}\,|\,\Dcal) \times \\
  \nonumber & \prod_{i=1}^{\nhat+1} P(c_{i} \,|\, s_{i}, t_{i}, \ab,
  \vb_{i}) \times P(t_{i}\,|\,c_{i-1}, \alphab^{\eta}) \times \\
  \nonumber & P(\vb_{i}\,|\,c_{i-1}, \Ab^{\gamma}) \times
  P(s_{i}\,|\,c_{i-1}, \Dcal) ,
\end{align}
where we define 
\begin{subequations}
  \label{eq:probcdef}
  \begin{align}
    \label{eq:aband1}
    P(c_i=0 \,|\, s_{i}=1) & = 1 \\
    \label{eq:aband2}
    P(c_i=0 \,|\, t_i=1) & =1 \\
    \label{eq:cnontrivial}
    P(c_i = j \,|\, s_{i},t_{i} = 0, \ab, \vb_{i}) & =
    P\rbr{a_j=1} \cdot P\rbr{v_{i,j}=1}. 
  \end{align}
\end{subequations}
If the user has abandoned the session ($s_{i}=1$ or $t_{i}=1$) then
subsequent value of $c_{i}$ is set to $0$. This is captured in
\eqref{eq:aband1} and \eqref{eq:aband2} above. On the other hand, if the
user decides to continue, then the probability of a click at the $j$-th
position (with $j \neq 0$) depends on the attractiveness of the document
as well as the position bias as can be seen from \eqref{eq:cnontrivial}.

\subsection{Inference}
\label{sec:Inference}

Global parameters of our model include $\eta_{i}$, and $\gammab_{i}$ for
$i=1,\ldots, n$. Furthermore, for each document $d$ we need to infer its
attractiveness $\theta_{d}$ and post-click relevance $\rho_{d}$. The
hyper-parameters $\alphab^{\eta}$ and $\Ab^{\gamma}$ are estimated from
historical data, and the other parameters such as $\eta_{i}$,
$\gammab_{i}$, and $\rho_{d}$ can be estimated efficiently by one pass
through the click-logs. Details can be found in
Appendix~\ref{sec:ParameterEstimation}. Next we discuss a novel method
for computing $\theta_{d}$.

\subsection{Estimating Attractiveness}
\label{sec:attmodel}

A sponsored search result $d$ displayed in response to a query $q$
contains three components namely a title, description, and a display
URL. We use the following naive Bayes like scheme to estimate
attractiveness: For each word $w$, excluding commonly occurring stop
words, which appears either in the title, description or the display
URL\footnote{For simplicity the display URL is treated as a single
  word.} we compute $N_{w}$ the number of occurrences of $w$ in
$\Dcal_{q}$, and $\hat{N}_{w}$ the number of occurrences of $w$ in the
subset of $\Dcal_{q}$ that received a click. For infrequent queries (search
volume less than 50 in a one week window) we let $\Dcal_{q}$ be the
entire training corpus, while for frequent queries (which occur with
frequency greater than 50) we restrict $\Dcal_{q}$ to the documents
displayed in response to $q$. Let $|d|$ denote the number of words in
$d$ and estimate
\begin{align}
  \theta_d = \frac{1}{\abr{d}} \sum_{\forall w \in d}
  \frac{\Nhat_{w}}{N_{w}}. 
\end{align}
In other words, for each word $w \in d$ we estimate the fraction of
times it occurred in a clicked document for query $q$, sum the
contribution due to each word independently and normalize by the length
of $d$ to obtain $\theta_d$. The advantage of our method is that we
share information across documents in the collection $\Dcal_{q}$. For
instance, if a word $w$ occurs frequently in documents which are clicked
in response to a query, then it is very likely to be a relevant word for
that query and our method gives it a high weight. Another advantage of
our method is that we are able to address the cold start problem; we can
estimate the attractiveness of a new sponsored search result based on
the words that appear in the ad copy.

\section{Experimental Evaluation}
\label{sec:ExperEval}

We collected three weeks of click logs from a large commercial search
engine, and used the first week data for computing priors, the second
week data for training, and third week data for testing. We filter the
data and only retain sessions where at least one mainline ad was
displayed.  Our test setup is closer to a real-world deployment
scenario, and somewhat different from the commonly used practice of
randomly splitting available data into a training and test set.  In
particular, we retain all queries in the test set, even if they have not
been observed in the training set. In this case, our estimation is
purely based on priors, which are computed from the first week of
data. We set $\alpha_i^\gamma + \beta_i^\gamma = 10$ and  $\alpha_i^\eta + \beta_i^\eta =10$.
Table~\ref{tab:datastats} summarizes our training data statistics. Note
that for high decile (>1000) queries, there are significantly fewer
three and four click sessions than for tail queries.

\subsection{Baselines}
\label{sec:Baselines}
For our experiments we will compare the performance of our algorithm
against the following four baselines:

\begin{enumerate}
\item Dynamic Bayesian Network (DBN) model: The user perseverance
  parameter in DBN (see \eqref{eq:dbn-model-examine}) was set to
  $\lambda_{i-1}=\lambda = 0.01$ for all $i$. This was found via
  cross-validation, based on the recommendation of \cite{ChaZha09}.
\item Independent Click Model (ICM): Since the cascade model
  \cite{CraZoeTayRam08} can only handle single click sessions, one can
  extended it by setting the user perseverance parameter $\lambda$ in
  \eqref{eq:dcm-model} to one \cite{GuoLiuWan09}. This yields the ICM. 
\item Position Model (PM): In this model, the click sequence only
  depends on the position of an ad, and is independent of its contents
  or attractiveness.  In other words, the first click is at position
  one, second click on position two and so on.  Because of position bias
  (see Table~\ref{tab:pos-click}), this is a strong baseline.
\item Attractiveness model (AM): The click sequence only depends on
  the attractiveness of an ad (section \ref{sec:attmodel}), and is
  independent of its position. In other words, the most attractive ad is
  clicked first and so on.
\end{enumerate}

\begin{table}[tbp]
  \caption{Because of confidentiality concerns we cannot report the size
    of our data set. Instead, we report data statistics which are normalized
    by the number of sessions in the 0-10 bucket.
    For each decile we also show
    the fraction of sessions which received one to four clicks.\newline
  }  
  \begin{tabular}[htbp]{|r|r|r|r|r|r|}
    \hline
    Freq.\ & Frac.\ & 1 Click & 2 Click & 3 Click & 4 Click\\
    \hline
    0-10 & 1.0 & 91.6\%& 7.3\%& 1\%& 0.1\%\\
    \hline
    10-50 & 0.39& 91.5\%& 7.3\%& 1.1\%&0.1\%\\
    \hline
    50-100 & 0.16& 92.3\%& 6.6\%& 1\%&0.1\%\\
    \hline
    100-500 & 0.37& 93.5\%& 5.7\%&0.7\% &0.1\%\\
    \hline
    500-1000 & 0.15&95.4\% & 4.1\%&0.42\% &0.08\%\\
    \hline
    >1000 & 1.0&97.94\% & 1.9\%&0.15\% &0.01\%\\
    \hline
  \end{tabular}
  \label{tab:datastats}
\end{table}

\begin{table}[tbp]
  \centering
  \caption{First click distribution across the four mainline positions. \newline}
  \begin{tabular}{cccc}
    \hline
    1 & 2 & 3 & 4\\	
    \hline
    0.708 & 0.163 & 0.0787 &	0.0503\\
    \hline
  \end{tabular}
  \label{tab:pos-click}
\end{table}
\subsection{Evaluation Plan}
\label{sec:EvaluationPlan}


In the first part of our experimental evaluation we are mainly
interested in understanding how well our algorithm is able to infer
post-click relevance. After all, this is the main motivation behind
studying click models. Since other models do not distinguish between
attractiveness and post-click relevance, we will only compare our
algorithm with the dynamic Bayesian network (DBN) model of
\cite{ChaZha09}. Details of this experiment can be found in
Section~\ref{sec:relad}.

For our second set of experiments (Sections \ref{sec:PredFirstClick} to
\ref{sec:PredReverseClick}) we focus on verifying that our model is able
to predict the sequence of user clicks accurately. Click models are
usually evaluated by computing average perplexity, or the closely
related log-likelihood on the test set. Recall that the perplexity for a
single user session is computed as
\begin{align}
  \label{eq:perplexity}
  p = 2^{-\frac{1}{n} \sum_{i=1}^{n} C_{i} \log q_{i} + \rbr{1-C_{i}}
    \log \rbr{1 - q_{i}}},
\end{align}
where $q_{i}$ is the probability of observing a click at position $i$ as
predicted by the model and $C_{i}$ indicates if an actual user click was
observed at that position.

However, we believe that perplexity alone does not tell the full
story. For instance, the perplexity scores of our model (1.1932) and DBN
(1.1984) are very similar, but they differ vastly in terms of how well
they predict a sequence of user clicks. Similarly, a model which simply
predicts $q_i$ as the empirical click-through rate, especially for high
decile queries, achieves very low perplexity but is unable to explain a
sequence of user clicks (also see the results for the AM
below). Therefore, we adapt a stronger evaluation criterion which is
designed to answer the following natural questions:
\begin{itemize}
\item How well does the model predict the first click position?
\item How well does the model predict two, three, and all four click
  sequences? This is a very stringent evaluation criterion for multiple
  click models, and the model wins only if it predicts all clicks in the
  sequence correctly.
\item Even if the model makes mistakes in predicting the actual click
  sequence, we want to understand whether the actual click sequence
  ranks high in terms of log-likelihood.
\item Does the model predict the top two and three clicks correctly?  In
  other words, the predicted click sequence need not be in the same
  order as the actual click sequence but the model needs to accurately
  predict whic h ads were clicked. For instance, if the actual click
  sequence was $\cbr{1,2}$ and the model predicts $\cbr{2,1}$ then this
  is considered to be a correct prediction as per this metric. Note that
  predicting the top four clicks in a four click session will always be
  100\% accurate.
\item How does the model fare in terms of predicting and ranking of
  sessions with reverse clicks?
\end{itemize}

Intuitively, predicting well on higher decile queries is easier than
lower decile queries. In order to understand the strengths and
weaknesses of various model, for all the above cases we report their
performance across different query deciles. 

\subsection{Predicting Post-Click Relevance}
\label{sec:relad}

\begin{figure}[htbp]
  \begin{center}
    \begin{tikzpicture}[scale=0.9] 
      \begin{axis}[
        xlabel=Recall, xlabel near ticks,
        ylabel=Precision, ylabel near ticks,
        ymin=0.7, ymax=1,
        xmin=0.001, xmax=1,
        legend pos=north east,]
        \addplot  table[x=recall,y=precision] {roc_1.txt};
        \addplot table[x=dbnrecall,y=dbnprecision] {roc_1.txt};        
        \legend{Our Model, DBN}
      \end{axis}
    \end{tikzpicture}
  \end{center}
  \caption{Area Under Curve for predicting relevance of Ads} 
  \label{fig:rel}
\end{figure}
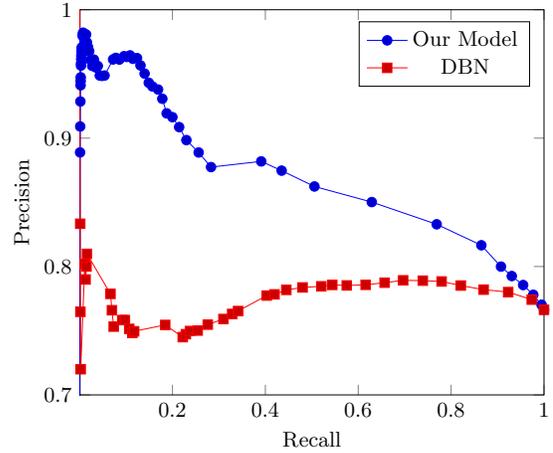

We have access to approximately 10,000 (query, ad) pairs from the test
dataset which have been labeled as relevant or irrelevant by trained
human editors. Note that the human editors look into the landing pages
to determine the labels. Following \cite{ChaZha09} we rank these $(q,
d)$ pairs using a score which is calculated as $\theta_{d} \times
\rho_{d}$ (attractiveness times post-click relevance). The documents are
ranked based on the computed score, and we measure precision vs
recall. The results for our model and DBN are plotted in
Figure~\ref{fig:rel}. Clearly our proposed model outperforms DBN
consistently across recalls. In particular note that our model has high
precision at low recall and is therefore able to rank relevant documents
higher on the list as compared to DBN. Our model achieves an AUC of
0.8653 compared to DBN which is only able to achieve 0.7843. 

\subsection{Predicting the First Clicked Ad}
\label{sec:PredFirstClick}

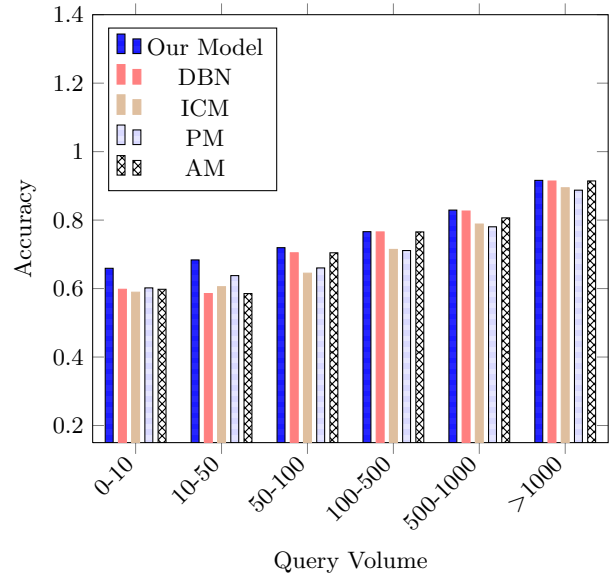
\begin{figure}[h]
  \begin{center}
    \begin{tikzpicture}[scale=1.0] 
      \begin{axis}[ybar,
        xlabel=Query Volume, xlabel near ticks,
        ylabel=Accuracy, ylabel near ticks,
        ymin=0.15, ymax=1.4,
        symbolic x coords={0-10,10-50,50-100,100-500,500-1000,>1000},
        xtick=data, x tick label style={rotate=45,anchor=east}, 
        legend pos=north west,
        bar width=3pt,]
        \addplot[pattern=horizontal lines dark blue] table[x=bin,y=tcm1] {forward_clicks.txt};
        \addplot[pattern=vertical lines, color=red!50] table[x=bin,y=dbn1] {forward_clicks.txt};
        \addplot[pattern=vertical lines, color=brown!50] table[x=bin,y=ccm1] {forward_clicks.txt};
        \addplot[pattern=horizontal lines light blue] table[x=bin,y=base1] {forward_clicks.txt};
        \addplot[pattern=crosshatch] table[x=bin,y=att_base1] {forward_clicks.txt};
        \legend{Our Model, DBN, ICM, PM, AM}
      \end{axis}
    \end{tikzpicture}
  \end{center}
  \caption{Accuracy of first click sequence prediction across different
    query deciles.} 
  \label{fig:firstclick}
\end{figure}

This is a multi-class classification problem with imbalanced class
probabilities. Therefore PM which predicts that the first click happens
at position one and has an accuracy of 72.27\% on our test dataset. The
AM has an accuracy of 73.5\% while the DBN and ICM accuracies are 72.7\%
and 71.7\% respectively.  In our model we predict the ad which has the
maximum click propensity as the first clicked ad. This results in an
accuracy of 79.59\%, a gain of over ~8.2\% as compared to the other
models.  To further understand the performance of models across
different query deciles we plot the accuracy of different models in
Figure \ref{fig:firstclick}. Note that we consistently outperform all
other models across all query deciles, with the gains being more
substantial for the lower decile queries which are harder to learn.

\begin{figure*}
  \begin{tikzpicture}[scale=0.7] 
    \begin{axis}[ybar,
      xlabel=Query Volume, xlabel near ticks,
      ylabel=Accuracy, ylabel near ticks,
      ymin=0.0, ymax=0.6,
      symbolic x coords={0-10,10-50,50-100,100-500,500-1000,>1000},
      xtick=data, x tick label style={rotate=45,anchor=east}, 
      legend pos=north west,
      bar width=4pt,]
      \addplot[pattern=horizontal lines dark blue] table[x=bin,y=tcm2] {forward_clicks.txt};
      \addplot[pattern=vertical lines, color=red!50] table[x=bin,y=dbn2] {forward_clicks.txt};
      \addplot[pattern=vertical lines, color=brown!50] table[x=bin,y=ccm2] {forward_clicks.txt};
      \addplot[pattern=horizontal lines light blue] table[x=bin,y=base2] {forward_clicks.txt};
      \addplot[pattern=crosshatch] table[x=bin,y=att_base2] {forward_clicks.txt};
      \legend{Our Model, DBN, ICM, PM ,AM}
    \end{axis}
  \end{tikzpicture}
  \begin{tikzpicture}[scale=0.7]
    \begin{axis}[ybar,
      xlabel=Query Volume, xlabel near ticks,
      ymin=0.0, ymax=0.7,
      symbolic x coords={0-10,10-50,50-100,100-500,500-1000,>1000},
      xtick=data, x tick label style={rotate=45,anchor=east}, 
      legend pos=north west,
      bar width=4pt,]
      \addplot[pattern=horizontal lines dark blue] table[x=bin,y=tcm3] {forward_clicks.txt};
      \addplot[pattern=vertical lines, color=red!50] table[x=bin,y=dbn3] {forward_clicks.txt};
      \addplot[pattern=vertical lines, color=brown!50] table[x=bin,y=ccm3] {forward_clicks.txt};
      \addplot[pattern=horizontal lines light blue] table[x=bin,y=base3] {forward_clicks.txt};
      \addplot[pattern=crosshatch] table[x=bin,y=att_base3] {forward_clicks.txt};
      \legend{Our Model, DBN, ICM, PM, AM}
    \end{axis}
  \end{tikzpicture}
  \begin{tikzpicture}[scale=0.7] 
    \begin{axis}[ybar,
      xlabel=Query Volume, xlabel near ticks,
      ymin=0.0, ymax=0.9,
      symbolic x coords={0-10,10-50,50-100,100-500,500-1000,>1000},
      xtick=data, x tick label style={rotate=45,anchor=east}, 
      legend pos=north west,
      bar width=4pt,]
      \addplot[pattern=horizontal lines dark blue] table[x=bin,y=tcm4] {forward_clicks.txt};
      \addplot[pattern=vertical lines, color=red!50] table[x=bin,y=dbn4] {forward_clicks.txt};
      \addplot[pattern=vertical lines, color=brown!50] table[x=bin,y=ccm4] {forward_clicks.txt};
      \addplot[pattern=horizontal lines light blue] table[x=bin,y=base4] {forward_clicks.txt};
      \addplot[pattern=crosshatch] table[x=bin,y=att_base4] {forward_clicks.txt};
      \legend{Our Model, DBN, ICM, PM, AM}
    \end{axis}
  \end{tikzpicture}
  \caption{Accuracy of two (L), three (M), four (R) click sequence
    prediction across different query deciles.}
  \label{fig:multiclick}
\end{figure*}
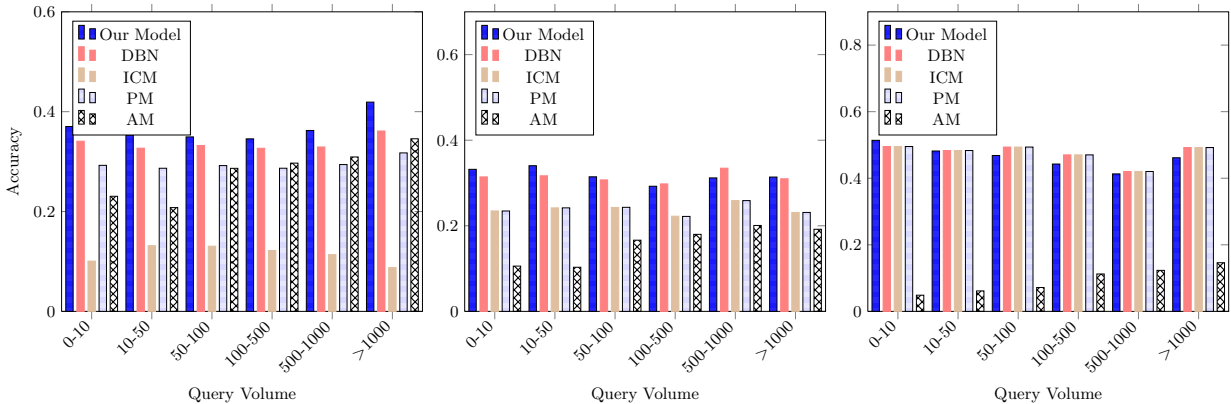

\subsection{Predicting the Entire Click Sequence}
\label{sec:PredEntirSequ}

We compute click propensity for each click sequence with the same length
as the actual click sequence. We say that our model predicted the click
sequence only when we predict the entire click sequence
correctly. Table~\ref{table:clickseq} summarizes the results. Our model
gains 8.6\% (resp.\ 25\%) on two-click-sequence prediction and 3.4\%
(resp.\ 37\%) on three-click-sequence prediction over DBN (resp.\ PM).
The model accuracies are comparable when predicting four click
sequences, which are only a very small fraction of the data.

Although AM is very competitive when predicting the first click, it is
unable to predict longer click sequences accurately. This is because
users do not decide to click on an ad solely based on its
attractiveness, position bias and past click experience also plays an
important role and needs to be taken into account. 

\begin{table}[h]
  \caption{Accuracy of Predicting the Entire Click Sequence.} 
  \label{table:clickseq}
  \centering  
  \begin{tabular}{c c c c c c}
    \hline 
    \# Clicks & Our Model & DBN & PM & AM & ICM\\
    \hline
    2 & 36.71 & 33.79 & 29.35  & 25.44 & 10.94\\
    3 & 32.26 & 31.27 & 23.54  & 12.80 & 23.55\\
    4 & 47.94 & 48.58 & 48.58 & 6.86 & 48.58\\
    \hline
  \end{tabular}
\end{table}

Figure~\ref{fig:multiclick} shows the accuracy of the models for two,
three and four click prediction across different query deciles. Note
that we consistently outperform other models across all query deciles in
two click sequence prediction. In three click sequence prediction, DBN
performs slightly better in top deciles than our model. Since number of
queries and session in tail deciles are higher than top deciles, we
achieve better overall performance than DBN.

\subsection{Ranking of the Actual Click Sequence}
\label{sec:RankingActualClick}

In the previous section we focused on predicting the entire click
sequence correctly. Here we focus on how we rank the actual click
sequences in terms of log-likelihood. For this, we compute the log
likelihood for all permutations of click sequences. We then sort the
click sequences based on the computed log likelihood and check the rank
of the actual click sequence in this list. Table \ref{table:rank}
summarizes our results. As can be seen, our model on the average ranks
the actual click sequence among the top 3 or 4 of all the possible
permutations of click sequence.

\begin{table}
  \caption{Overall Actual Click Sequence Ranking.} 
  \label{table:rank}
  \centering  
  \begin{tabular}{c c}
    \hline
    Number of Clicks & Average Rank \\
    \hline 
    1 & 1.25   $\pm$ 0.62\\
    2 &  2.25  $\pm$ 2.27\\
    3 &  2.67  $\pm$ 4.17\\
    4 &  1.87  $\pm$ 3.36\\
    \hline 
  \end{tabular}
\end{table}

Figure~\ref{fig:rank} shows how the ranking of the actual click sequence
varies across different query deciles. As can be seen from
Table~\ref{tab:datastats}, longer click sequences are more frequent in
the lower deciles, and hence our model has more data to learn in these
deciles. Consequently the average rank of the actual click sequence as
predicted by our model for two, three and four clicks is lower for lower
decile queries.

\begin{figure}[htbp]
  \begin{center}
    \begin{tikzpicture}[scale=1.0] 
      \begin{axis}[ybar,
        xlabel=Query Volume, xlabel near ticks,
        ylabel=Rank of Actual Click Sequence, ylabel near ticks,
        ymin=0.0,ymax=11.0,
        symbolic x coords={0-10,10-50,50-100,100-500,500-1000,>1000},
        xtick=data,x tick label style={rotate=45,anchor=east}, 
        bar width=5pt,
        legend style={at={(0.5,1.15)},
          anchor=north,legend columns=-1},
        error bars/y dir=both, 
        error bars/y explicit,
        error bars/error bar style={black,line width=0.5pt},
        ]
        \addplot[pattern=horizontal lines light blue] table[x=bin,y=rank1, y error=stdrank1] {forward_clicks.txt};
        \addplot[pattern=horizontal lines dark blue] table[x=bin,y=rank2, y error=stdrank2] {forward_clicks.txt};
        \addplot[pattern=horizontal lines, color=red!30] table[x=bin,y=rank3, y error=stdrank3] {forward_clicks.txt};
        \addplot[pattern=vertical lines, color=green!50] table[x=bin,y=rank4, y error=stdrank4] {forward_clicks.txt};
        \legend{Single Click, Two Click, Three Click, Four Click}
      \end{axis}
    \end{tikzpicture}
  \end{center}
  \caption{Actual click sequence ranking across different query deciles.} 
  \label{fig:rank}
\end{figure}

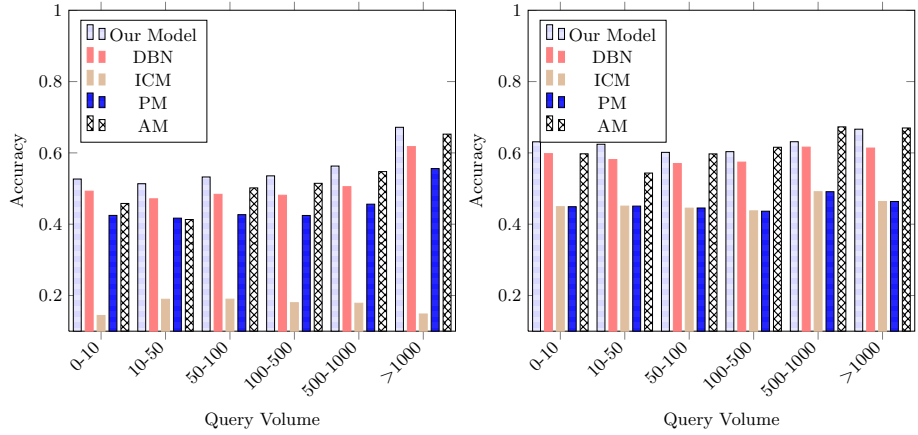
\begin{figure*}
  \centering
  \begin{tikzpicture}[scale=0.75] 
    \begin{axis}[ybar,
      xlabel=Query Volume, xlabel near ticks,
      ylabel=Accuracy, ylabel near ticks,
      ymin=0.1,ymax=1,
      symbolic x coords={0-10,10-50,50-100,100-500,500-1000,>1000},
      xtick=data,x tick label style={rotate=45,anchor=east}, 
      legend pos=north west,
      bar width=4pt,]
      \addplot[pattern=horizontal lines light blue] table[x=bin,y=topN2] {forward_clicks.txt};
      \addplot[pattern=vertical lines, color=red!50] table[x=bin,y=dbn_topN2] {forward_clicks.txt};
      \addplot[pattern=vertical lines, color=brown!50] table[x=bin,y=ccm_topN2] {forward_clicks.txt};
      \addplot[pattern=horizontal lines dark blue] table[x=bin,y=baseonly_topN2] {forward_clicks.txt};
      \addplot[pattern=crosshatch] table[x=bin,y=attonly_topN2] {forward_clicks.txt};
      \legend{Our Model, DBN, ICM, PM, AM}
    \end{axis}
  \end{tikzpicture}
  \begin{tikzpicture}[scale=0.75] 
    \begin{axis}[ybar,
      xlabel=Query Volume, xlabel near ticks,
      ylabel=Accuracy, ylabel near ticks,
      ymin=0.1,ymax=1,
      symbolic x coords={0-10,10-50,50-100,100-500,500-1000,>1000},
      xtick=data,x tick label style={rotate=45,anchor=east}, 
      legend pos=north west,
      bar width=4pt,]
      \addplot[pattern=horizontal lines light blue] table[x=bin,y=topN3] {forward_clicks.txt};
      \addplot[pattern=vertical lines, color=red!50] table[x=bin,y=dbn_topN3] {forward_clicks.txt};
      \addplot[pattern=vertical lines, color=brown!50] table[x=bin,y=ccm_topN3] {forward_clicks.txt};
      \addplot[pattern=horizontal lines dark blue] table[x=bin,y=baseonly_topN3] {forward_clicks.txt};
      \addplot[pattern=crosshatch] table[x=bin,y=attonly_topN3] {forward_clicks.txt};
      \legend{Our Model, DBN, ICM, PM, AM}
    \end{axis}
  \end{tikzpicture}
  \caption{Accuracy of predicting positions of two (L) and three (R)
    click sessions across different query deciles.}
  \label{fig:topclicks}
\end{figure*}
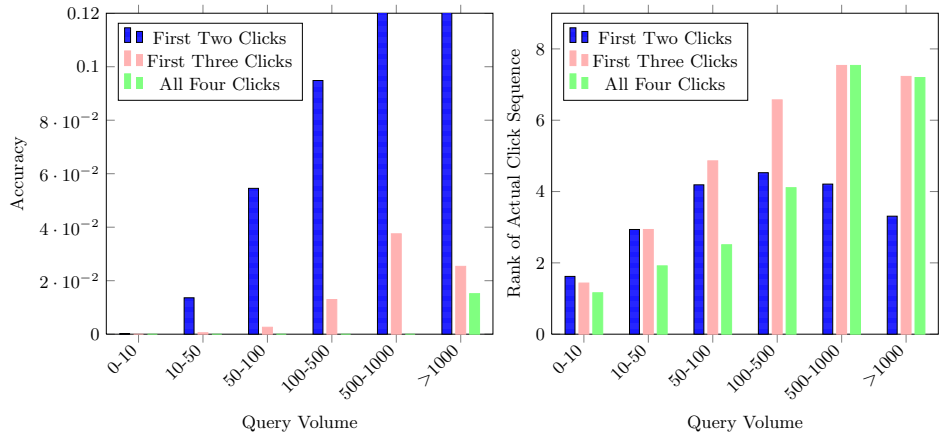
\begin{figure*}
  \begin{center}
    \begin{tikzpicture}[scale=0.75] 
      \begin{axis}[ybar,
        xlabel=Query Volume, xlabel near ticks,
        ylabel=Accuracy, ylabel near ticks,
        ymin=0.0,ymax=0.12,
        symbolic x coords={0-10,10-50,50-100,100-500,500-1000,>1000},
        xtick=data,x tick label style={rotate=45,anchor=east}, 
        legend pos=north west,
        bar width=5pt,]
        \addplot[pattern=horizontal lines dark blue] table[x=bin,y=tcm2] {reverse_clicks.txt};
        \addplot[pattern=horizontal lines, color=red!30] table[x=bin,y=tcm3] {reverse_clicks.txt};
        \addplot[pattern=vertical lines, color=green!50] table[x=bin,y=tcm4] {reverse_clicks.txt};
        \legend{First Two Clicks, First Three Clicks, All Four Clicks}
      \end{axis}
    \end{tikzpicture}
    \begin{tikzpicture}[scale=0.75] 
      \begin{axis}[ybar,
        xlabel=Query Volume, xlabel near ticks,
        ylabel=Rank of Actual Click Sequence, ylabel near ticks,
        ymin=0.0,ymax=9.0,
        symbolic x coords={0-10,10-50,50-100,100-500,500-1000,>1000},
        xtick=data,x tick label style={rotate=45,anchor=east}, 
        legend pos=north west,
        bar width=5pt,]
        \addplot[pattern=horizontal lines dark blue] table[x=bin,y=rank2] {reverse_clicks.txt};
        \addplot[pattern=horizontal lines, color=red!30] table[x=bin,y=rank3] {reverse_clicks.txt};
        \addplot[pattern=vertical lines, color=green!50] table[x=bin,y=rank4] {reverse_clicks.txt};
        \legend{First Two Clicks, First Three Clicks, All Four Clicks}
      \end{axis}
    \end{tikzpicture}
  \end{center}
  \caption{Accuracy (L) of predicting the entire click sequence, and
    ranking (R) of the actual click sequence  across different query
    deciles for sessions where at least one pair of clicks was observed
    in reverse order.} 
  \label{fig:reverse}
\end{figure*}

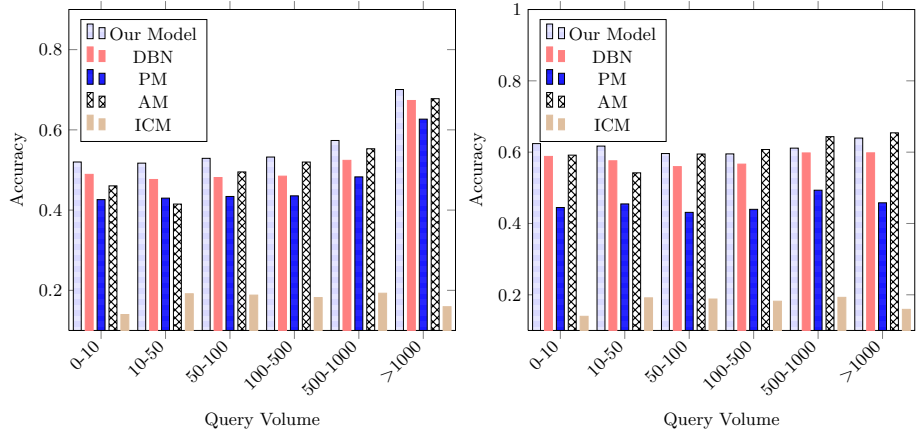
\begin{figure*}
  \begin{center}
    \begin{tikzpicture}[scale=0.750] 
      \begin{axis}[ybar,
        xlabel=Query Volume, xlabel near ticks,
        ylabel=Accuracy, ylabel near ticks,
        ymin=0.1,ymax=0.9,
        symbolic x coords={0-10,10-50,50-100,100-500,500-1000,>1000},
        xtick=data,x tick label style={rotate=45,anchor=east}, 
        legend pos=north west,
        bar width=4pt,]
        \addplot[pattern=horizontal lines light blue] table[x=bin,y=topN2] {reverse_clicks.txt};
        \addplot[pattern=vertical lines, color=red!50] table[x=bin,y=dbn_topN2] {reverse_clicks.txt};
        \addplot[pattern=horizontal lines dark blue] table[x=bin,y=baseonly_topN2] {reverse_clicks.txt};
        \addplot[pattern=crosshatch] table[x=bin,y=attonly_topN2] {reverse_clicks.txt};
        \addplot[pattern=vertical lines, color=brown!50] table[x=bin,y=ccm_topN2] {reverse_clicks.txt};
        \legend{Our Model, DBN, PM, AM, ICM}
      \end{axis}
    \end{tikzpicture}
    \begin{tikzpicture}[scale=0.75] 
      \begin{axis}[ybar,
        xlabel=Query Volume, xlabel near ticks,
        ylabel=Accuracy, ylabel near ticks,
        ymin=0.1,ymax=1,
        symbolic x coords={0-10,10-50,50-100,100-500,500-1000,>1000},
        xtick=data,x tick label style={rotate=45,anchor=east}, 
        legend pos=north west,
        bar width=4pt,]
        \addplot[pattern=horizontal lines light blue] table[x=bin,y=topN3] {reverse_clicks.txt};
        \addplot[pattern=vertical lines, color=red!50] table[x=bin,y=dbn_topN3] {reverse_clicks.txt};
        \addplot[pattern=horizontal lines dark blue] table[x=bin,y=baseonly_topN3] {reverse_clicks.txt};
        \addplot[pattern=crosshatch] table[x=bin,y=attonly_topN3] {reverse_clicks.txt};
        \addplot[pattern=vertical lines, color=brown!50] table[x=bin,y=ccm_topN2] {reverse_clicks.txt};
        \legend{Our Model, DBN, PM, AM, ICM}
      \end{axis}
    \end{tikzpicture}
  \end{center}
  \caption{Accuracy of predicting positions of two (L) and three (R)
    click sessions across different query deciles for sessions where at
    least one pair of clicks was observed in reverse order. }
  \label{fig:reverseclicks}
\end{figure*}

\subsection{Predicting Top Clicks}
\label{sec:PredictingTopClicks}

Here we ignore the order and focus on understanding if our model is able
to predict the \emph{positions} of the clicks in two and three click
sessions. Table~\ref{table:topn} shows we out perform all other models
in predicting top 2 and 3 click positions. Figure~\ref{fig:topclicks}
shows accuracies across query deciles. 

\begin{table}[h]
  \caption{Accuracy of Predicting the Locations of Top Clicks.}
  \label{table:topn}
  \centering  
  \begin{tabular}{c c c c c c}
    \hline
    \# Clicks & Our Model &  DBN & PM & AM & ICM \\ 
    \hline
    2 & 54.93 & 51.11 & 43.89 & 48.5 & 16.07\\
    3 &  62.46  & 59.14 & 44.98 & 59.7 & 45.01\\
    \hline 
  \end{tabular}
\end{table}

\subsection{Predicting Reverse Click Sequences}
\label{sec:PredReverseClick}

In our final experiment we focus only on the sessions where at least one
pair of clicks was observed in reverse order (reverse click
sessions). As before, we focus on the accuracy
(Table~\ref{table:reverse}), rank of the actual click sequence
(Table~\ref{table:reverserank}), and accuracy of predicting the location
of the top clicks
(Table~\ref{table:topnreverse}). Figure~\ref{fig:reverse} shows the
accuracy and rank of the actual click sequence for different deciles,
and Figure~\ref{fig:reverseclicks} shows the accuracy of predicting
positions of clicks in two and three click sessions.

As expected our model prediction accuracy for reverse click sessions is
low since the reverse click sessions are only around 30\% of the
multi-click sessions. Even though our accuracies are lower than the
attractiveness model, our model ranks actual reverse click sequences in
top 5 of all the possible permutations of click sequence. Our model out
performs all other models in most of the experiments and still achieves
better ranking of actual click sequence even when clicks have been
observed in reverse order.

\begin{table}
  \caption{Accuracy of Predicting the Entire Click Sequence when Actual
    Click Sequence has Reverse Order.} 
  \label{table:reverse}
  \centering  
  \begin{tabular}{c c c c c}
    \hline
    \# Clicks & Our Model & DBN & PM & AM \\
    \hline
    2 & 8.1 & 0.0 & 0.0 & 17.70\\
    3 & 0.6 & 0.0 & 0.0 & 9.4\\
    4 & 0.1 & 0.0 & 0.0 & 3.9\\
    \hline
  \end{tabular}
\end{table}

\begin{table}
  \caption{Overall Reverse Click Sequence Ranking.} 
  \label{table:reverserank}
  \centering  
  \begin{tabular}{c c c c}
    \hline
    Number of Clicks &  Average Rank\\
    \hline 
    2 & 3.15$\pm$3.01 \\
    3 & 3.89$\pm$5.46 \\
    4 & 2.67$\pm$4.56\\
    \hline
  \end{tabular}
\end{table}

\begin{table}
  \caption{Accuracy of Predicting Locations of Top Clicks when Actual Click
    Sequence has Reverse Order.} 
  \label{table:topnreverse}
  \centering  
  \begin{tabular}{c c c c c c}
    \hline
    \# Clicks & Our Model & DBN & PM & AM & ICM\\
    \hline
    2 & 55.74  & 51.12 & 45.66 & 49.50 & 16.09\\
    3 &  61.51 & 58.20 & 44.74 & 59.14 & 44.75\\
    \hline 
  \end{tabular}
\end{table}

\section{Conclusion and Future Work}
\label{sec:Conclusion}

We presented a new multiple click model for modeling how users interact
with and click on sponsored search results. Our model can handle reverse
click sequence and comprehensively outperforms other models across a
number of different metrics in extensive empirical evaluation.

Online sponsored search auctions are priced using a Generalized Second
Price (GSP) auction mechanism. Inherent in this model is the assumption
that the clicks on the ads happen independent of each other. We are
currently working on developing pricing mechanisms which will take into
account the clicking behavior predicted by our model. Our efforts are
also directed towards improving the reverse click prediction accuracy of
our model by using stronger priors. Finally, we are also working towards
making our model robust to noise.

\appendix

\section{Parameter Estimation}
\label{sec:ParameterEstimation}



Our training data consists of $m$ sessions, and we assume that the
document set $\Dcal^{k}$ was displayed in the $k$-th session in response
to query $q^{k}$ and we observed a click sequence $\cbb^{k}$ of length
$\nhat^{k}$. If we assume that the sessions are iid, then 
\begin{align}
  \label{eq:full-data}
  P\rbr{\cbr{\cbb^{k}} \,|\,\cbr{\Dcal^{k}}} & = \prod_{k=1}^{m}
  P\rbr{\cbb^{k} \,|\, \Dcal^{k}}. 
\end{align}
Let $c_{i}^{k}$ denote the location of the $i$-th click in the $k$-th
session, $I(\cdot)$ denote the indicator variable of an event, $\Dcal'$
denote the set of unique query-document pairs in our session data and $D
= \abr{\Dcal'}$. Plugging in \eqref{eq:ClickModel} into
\eqref{eq:full-data}, using \eqref{eq:probsdef}--\eqref{eq:probcdef}
shows that $P\rbr{\cbr{\cbb^{k}} \,|\,\cbr{\Dcal^{k}}}$
\begin{align}
  \nonumber 
  & \propto \prod_{d \in \Dcal'} \theta_{d}^{\psi_{d}} \  
  \times \prod_{j=1}^{n}\prod_{k=1}^{n} \gamma_{jk}^{\delta_{jk}} \, p\rbr{\gamma_{jk}} \\
  \nonumber & \times \prod_{j=0}^{n} \rbr{1-\eta_{j}}^{\beta_{j}}
  \eta_{j}^{\beta'_{j}} \, p\rbr{\eta_{j}} 
  \times \prod_{d \in \Dcal'} \rbr{1 - \rho_{d}}^{\kappa_{d}}
  \rho_{d}^{\kappa'_{d}},
\end{align}
where we define
\begin{align} 
  \label{eq:psi-def}
  \psi_{d} & = \sum_{k=1}^{m} \sum_{i=1}^{\nhat_{k}}
  I\rbr{d_{c_{i}^{k}} = d}\\
  \label{eq:delta-def}
  \delta_{i,j} & = \sum_{k=1}^{m} \sum_{t=1}^{\nhat_{k}} I\rbr{c_{t-1}^{k}
    = i} \text { and } I\rbr{c_{t}^{k} = j}. 
\end{align}
$\psi_{d}$ counts the number of times document $d$ occurs in
$\cbr{\Dcal_{k}}$, $\delta_{i,j}$ counts the number of times we observed
a click at position $j$ after having observed a click at position
$i$. Note that when $i < j$ $\delta_{i,j}$ counts the in-sequence users
clicks, and for all $i > j$, $\delta_{i,j}$ captures out of sequence
clicks.

\begin{align}
  \label{eq:beta-def}
  \beta_{j} = \sum_{k=1}^{m} I\rbr{\nhat^{k} > j}\ \textrm{and} \ 
  \beta'_{j} = \sum_{k=1}^{m} I\rbr{\nhat^{k} = j}\\
  \label{eq:betap-def}
\end{align}
$\beta_{j}$ counts the number of sessions with greater than $j$ clicks,
while $\beta'_{j}$ counts the number of sessions with exactly $j$
clicks. It is easy to see that $\beta_{j} = \sum_{j'>j} \beta'_{j'}$ for
$j\neq 0$.

\begin{align}
  \label{eq:kappa-def}
  \kappa_{d} = \sum_{k=1}^{m} I\rbr{d_{c^{k}_{\nhat^{k}}} \neq d} \text {
    and } I\rbr{d_{c_{i}^{k}} = d} \text{ for some } i, 
\end{align}
\begin{align}
  \label{eq:kappa-def}
  \text{and }  \kappa'_{d} = \sum_{k=1}^{m} I\rbr{d_{c^{k}_{\nhat^{k}}} = d}.
\end{align}
$\kappa'_{d}$ counts the number of sessions which ended after clicking
on the document $d$, while $\kappa_{d}$ is the number of sessions which
did not end after a click on document $d$. 

As a consequence of using the Beta prior, we can estimate
\begin{align}
  \label{eq:eta-est} 
  \eta_{j}=\frac{\beta'_j+\alpha_{j}^{\eta}}{\beta'_j+\alpha_{j}^{\eta}+
    \beta_j + \beta_{j}^{\eta}} \ \textrm{and} \ \rho_{d}=\frac{\kappa'_{d}}{\kappa'_{d}+\kappa_{d}} 
\end{align}
Furthermore,
since $\gammab_{i} \sim Dir(\alphab_{i}^\gamma)$, we can estimate:
\begin{align}
  \label{eq:gamma-est}
  \gamma_{i,j}=\frac{\delta_{i,j}+\alpha_{i,j}^{\gamma}}{\sum_j
    (\delta_{i,j}+\alpha_{i,j}^{\gamma})}. 
\end{align}

\subsection{Time Complexity}

In order to perform the updates \eqref{eq:eta-est} 
and \eqref{eq:gamma-est} we need to compute the quantities defined in
\eqref{eq:psi-def} -- \eqref{eq:kappa-def}. However, all these
quantities involve simple counts, which can be computed by using 
one-pass through the click logs. Therefore, we can conclude that
inference in our model is extremely salable. 


\bibliography{bibfile}
\end{document}